# Estimation of available bandwidth and measurement infrastructure for Russian segment of Internet

**Platonov A.P.[1), Sidelnikov D.I.[2), Strizhov M.V.[3), Sukhov A.M.[4)**

[1) *Russian Institute for Public Networks, Moscow, Russia, e-mail: plat@ripn.net*
[2) *Institute of Organic Chemistry of RAS, Moscow, Russia, e-mail: sid@freenet.ru*
[3) *Samara State Aerospace University, Samara, Russia, e-mail: strizhov@ip4tv.ru*
[4) *Internet TV ltd., Samara, Russia, e-mail: amskh@yandex.ru*

## Abstracts

In paper the method for estimation of available bandwidth is supposed which does not demand the advanced utilities. Our method is based on the measurement of network delay $D$ for packets of different sizes $W$. The simple expression for available bandwidth $B_{av} = (W_2 - W_1)/(D_2 - D_1)$ is substantiated. For the experimental testing the measurement infrastructure for Russian segment of Internet was installed in framework of RFBR grant 06-07-89074.

C 2.3 Network Operations, Network monitoring

## Introduction

In order to construct the fullest picture of a global network, it's monitoring, search of bottlenecks, and also development of the standards describing new appendices, the modern measuring infrastructure should be installed.

In Russia there are some measuring points of different measurement projects in the area of networking, for example PingER [11] in Institute of Theoretical and Experimental Physics (ITEP), but full access to the collected data is limited for researchers. Unfortunately, the current measurement points do not reflect structure of the Russian segment of a global network.

The available bandwidth [6,9] of a network path is an important performance metric and its end-to-end estimation has recently received significant attention. The available bandwidth is the maximum throughput that the path can provide to an application, given the path's current cross traffic load. Measuring available bandwidth is not only for knowing the network status, but also to provide information to network applications on how to control their outgoing traffic and fairly share the network bandwidth.

Bulk Transport Capacity (BTC) is a measure of a network's ability to transfer significant quantities of data with a single congestion - aware transport connection (e.g., TCP). The intuitive definition of BTC is the expected long-term average data rate (bits per second) of a single ideal TCP implementation over the path in question.

Measuring the available bandwidth is of great importance for predicting the end-to-end performance of applications, for dynamic path selection and traffic engineering, and for selecting between a numbers of differentiated classes of service. The available bandwidth is an important metric for several applications, such as grid, video and voice streaming, overlay routing, p2p file transfers, server selection, and interdomain path monitoring.

In order to measure different capacity metrics the installation of special utilities [8] is demanded at both ends of path. This is uncomfortable process especially for usual network users who try to install modern application like videoconference.

At present time powerful measurement system like RIPE Test Box is expanded [5]. Unfortunately, this system doesn't measure the available bandwidth but it collects the numerical values characterized the network heals like delay, jitter, routing path, etc. This data allows us to investigate the basic interdependences of available bandwidth from basic network parameters. Our aim is to estimate the available bandwidth from the delay value received from one point of path.

**Model**

J. Padhye V. Firoiu D. Towsley and J. Kurose [1,10] find that TCP throughput may be calculated as

$$B_{av} = \frac{W}{D} \qquad (1)$$

This formula may be used for calculation of available bandwidth between two network points that are connected immediately. Here $W$ is the size of transmitted packet and $D$ is the packet delay. The delay value is caused by such constant network factors as propagation delay, transmission delay, per-packet router processing time, etc [2].

Choi et al [2] proved that the fixed delay component (i.e., the total propagation and transmission delay) for a packet of size $p$ is a *linear* (or precisely, an *affine*) function of its size. To validate this assumption, they check the minimum delay of packets of the same size for each path, and plot the minimum delay against the packet size. But the parameters of their linear equation will not be a simple function of the link capacities and the propagation delays.

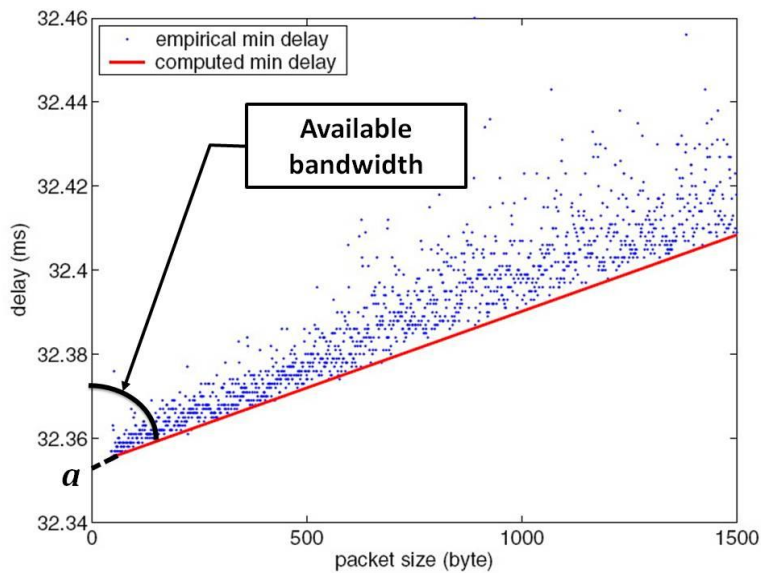

Fig.1. Available bandwidth

On the Fig.1 the graphic of linear dependence between network delay and packet size constructed in paper [2] shows computed minimal delay. Slope angle concerning Y axe could be considered as available bandwidth. Prolongation of this line to Y axe gives the intercept value a.

Then the equation (1) for the available bandwidth which path consists of two or more hops should be modernized to the following view:

$$B_{av} = \frac{W}{D-a} \qquad (2)$$

where $a$ is the unknown function depended on $n$ and $l$. Here $n$ is the number of hops (routers) that gives the *traceroute* command and $l = \sum_n l_n$ is the sum of single length of routing path. The value $D$ is related to the distance between the sites plus the propagation delay and per-packet router processing time at each hop along the path between the sites [3,4].

It is easy to use the linear approach for calculation of $a$

$$a = f(n,l) \approx \alpha n + \beta l \qquad (3)$$

In any case the equation (2) gives us the simple way for estimation of active bandwidth.

Our model supposes the variation of packet size on the same path for calculation of available bandwidth. If the testing process between two fixed points is organized by packets with different sizes $W_1$ and $W_2$ and then the $D$ times get also two different values. Equation (2) gives the identical value of available bandwidth independently from packet size. The system from two equations is easy solved to find $B_{av}$ and $a$ :

$$B_{av} = \frac{W_1 - W_2}{D_1 - D_2} \qquad (4)$$

The sizes $W_1$ and $W_2$ should different in several times, it is reasonable to choose 100 and 1124 *bytes* correspondingly. The basic problem is the precise of measurements for delay **D** that is necessary for accurate result. The delay should be measured with micro second precision, such accuracy could be reached from RIPE Test Box database.

Routinely the special utilities could be used for delay measurements; we tried to test traditional *ping,* the new *UDPping* and other utility. In result of test the simplest utility *ping* was found to be a best choice for delay measurements.

For example, ADSL connection at my home gives $D_1 = 18\ ms, D_2 = 42\ ms$, that corresponds to 350 *Kbps* of available bandwidth. During FTP session the delay grows to 300 *ms* and 425 *ms* that corresponds approximately to 60 *Kbps* of available bandwidth. This is very rough computation, but it could be made quickly and independently.

It should be noted that Table II from paper [2] gives us these values; calculated slope is inverse value to available bandwidth. The corresponding bandwidths for data set 1, 2, 3 (path 1 and 2) are 285 *Mbps*, 128 *Mbps*, 222 *Mbps* and 205 *Mbps*.

The second test allows calculating the value $a$ in alternative way.

$$a = \frac{W_2 D_1 - W_1 D_2}{W_2 - W_1} \qquad (6)$$

The further data from RIPE test box database allow calculating the coefficients from Eq.3.

The additional experiments on the different direction are necessary for verifying our model. During experiment the received data should be compared with results of *iperf* utility from NLANR.org. We have contacted with Baek-Young Choi from University of Minnesota. It will be interesting to test remote direction like Australia and New Zeland. Unfortunately, there is not a simple way to receive data with delays for different packet sizes in current configuration of RIPE Test Box.

**Measurement infrastructure**

In 2006 our team develops two stationary measuring points in the FREENet network on the basis of Institute of Organic Chemistry of the Russian Academy of Science and in the Samara regional network for a science and education at the Samara State Aerospace University within the limits of the RFBR grant 06-07-89074 "Creation of a measuring infrastructure for studying quality the Internet of appendices in the Russian segment of a global network".

Each of our point has two blocks:

1. Ripe Test Box [12]
2. Application analysis server (with tools such as Wireshark, H.323 Beacon Tool [7], etc.).

In December 2007 RIPE measuring system totaled about 80 points scattered worldwide. Practically all the cores world the Internet the centers are captured by system, data for the analysis the Internet of an infrastructure act from Northern and South America, Australia, New Zealand and Japan. But the greatest quantity of measuring points, as well as it is necessary to expect, is in a zone of responsibility RIPE: the European countries (Fig. 2).

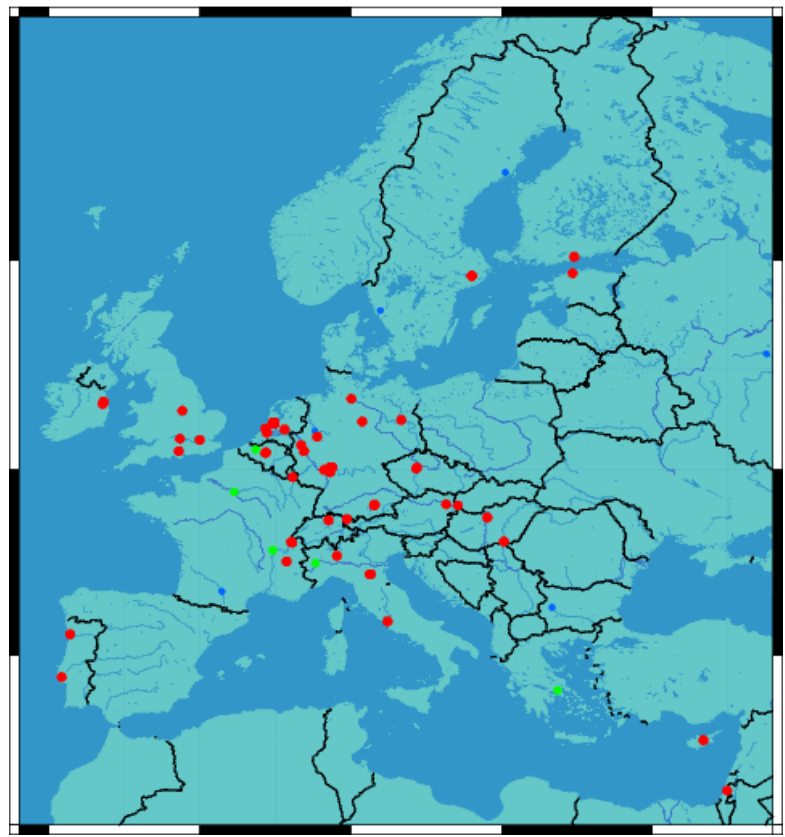

Fig 2. RIPE Test Box in Europe

Measured values includes:

1. Delay of packages
2. Variation of a delay (jitter)
3. Losses of packages
4. Availability of DNS root servers
5. Statistics of routing map
6. Availability of satellites

All these data are reduced together and results of their processing are accessible in the form of various schedules and tables. Work with data is convenient for beginning with starting page on the Internet to the address: http://ripe.net/projects/ttm/Plots/. Each of Test Boxes has its own RIPE identification (for example, Test box in Samara called tt143.ripe.net). Basis for the analysis is the summary table where data on a studied measuring point are together shown all, for example, tt143 is a point in Samara. In the table data about a time delay of packages are presented average percent of losses of packages in entering and proceeding traffic, and also: average size, the boundary sizes cutting first and last 2,5 % of results.

The data presented in the table, can be deciphered in more detail. Following a hyperlink in a corresponding cell of the table, it is possible to receive the additional information on dynamics of any of the measured sizes resulted above. So in a Fig. 3 presented data on parameter of a delay of packages between measuring points in Samara and Zurich, collected during between December 12th, 2007 and on January 12th, 2008 are presented.

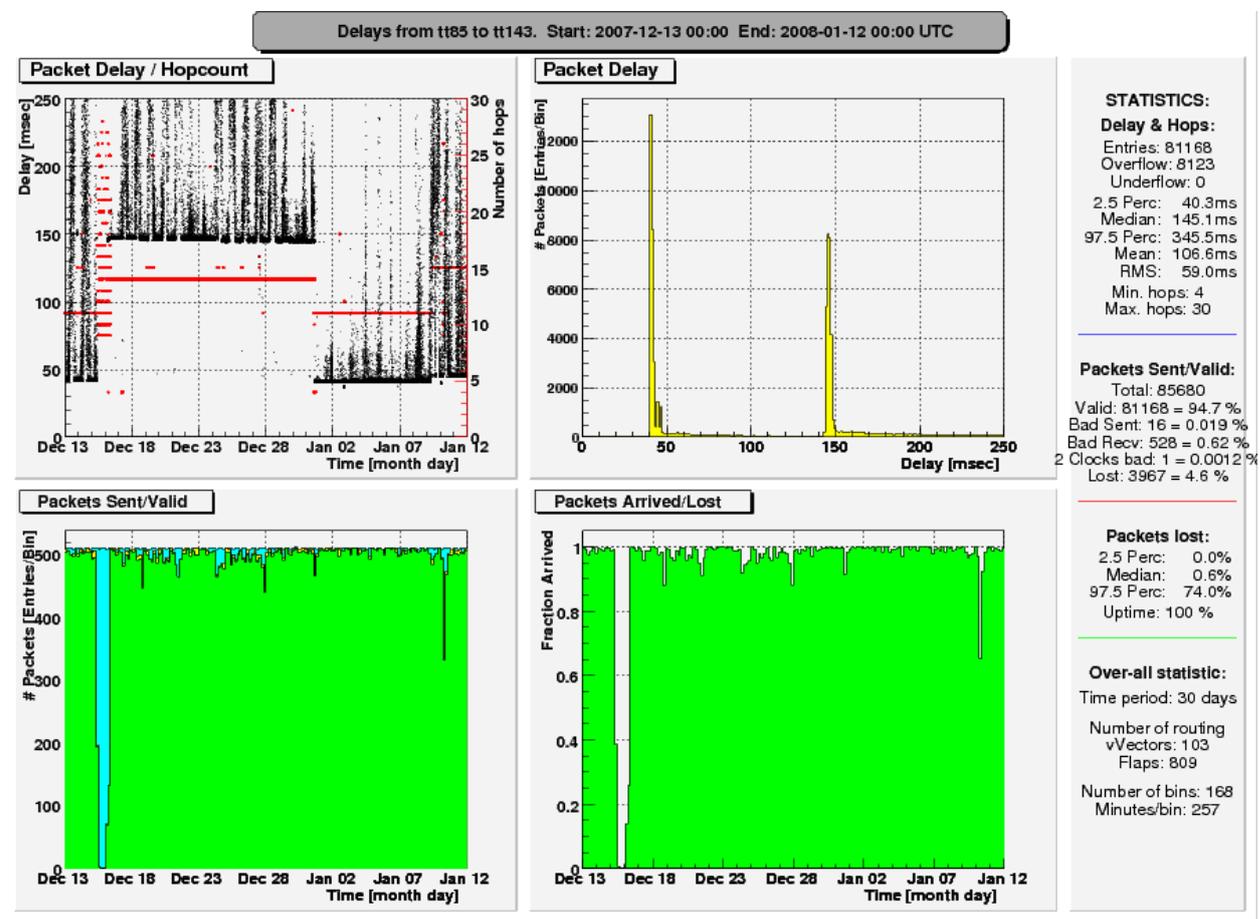

Fig. 3. Month statistics for delay

On the top left schedule data about two variables are presented:

1. To delay of packages depending on time
2. Routing hops

The presented schedule evidently shows that during supervision the route changed.

The top right schedule shows density of distribution for a delay of packages, at change of routing time of passage of a software package of a network tripled. Also the bottom right schedule showing losses of packages by transfer which allocates two time periods with practically full termination of communication is interesting also. The variation of a delay is size which dynamics of change measuring system RIPE allows to analyze (Fig. 4).

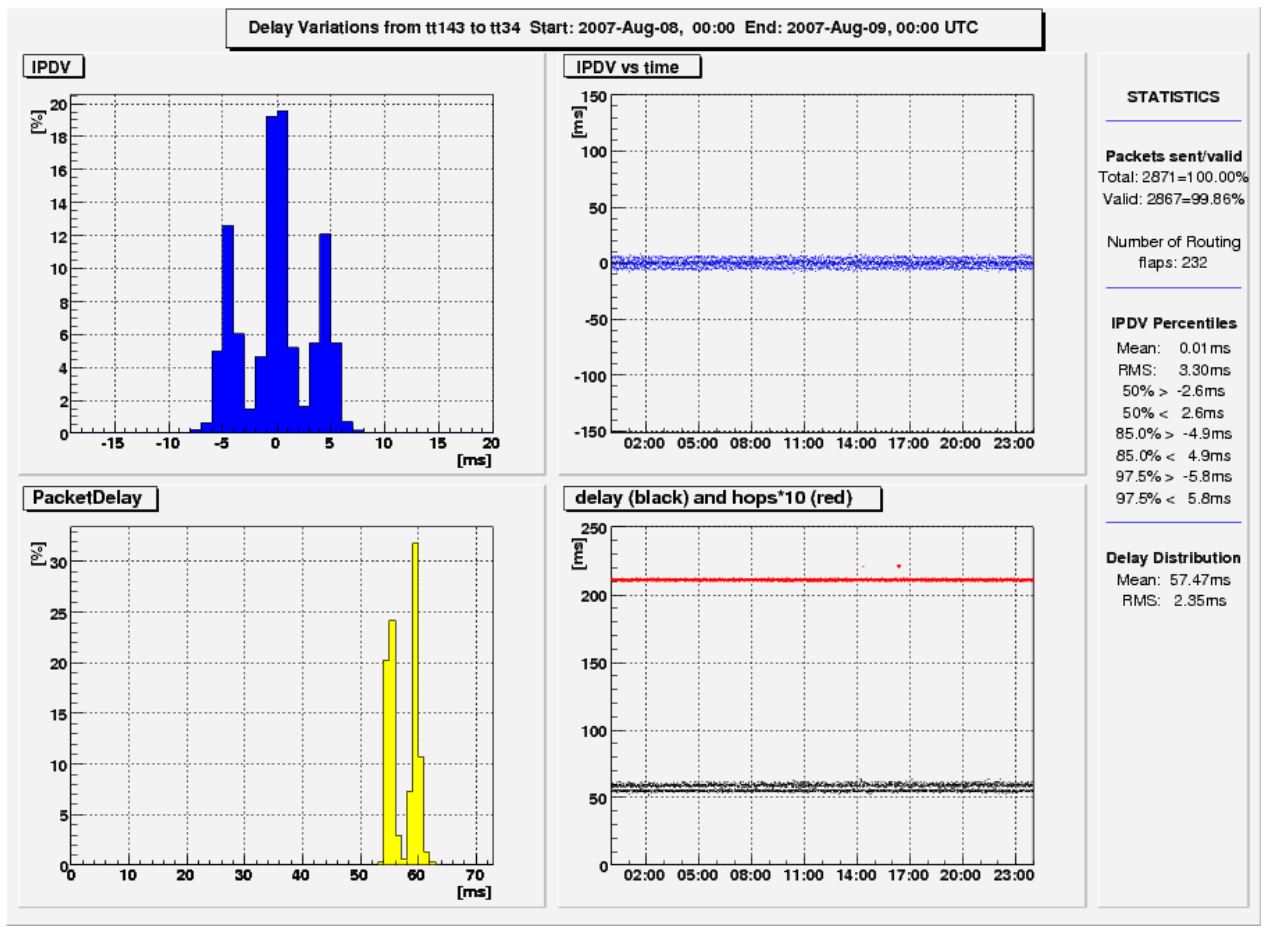

Fig. 4. Day statistics for packet loss

The top left schedule describes density of distribution for a variation of a delay, on right - time dependence of distribution for a variation of a delay.

On the bottom pair schedules the density of distribution and time dependence for a delay of packages, and also dynamics of change of number of points of routing are presented.

Especially it would be desirable to note the service showing availability root DNS of servers of domains of the first level. The given service automatically traces responses from almost 200 servers for 30 domains worldwide and conducts statistics of their availability. The specified service can be applied to monitoring national DNS systems, for this purpose it is enough to deduce of all data about availability of this or that server in a separate page.

The statistics of routing includes not only records of ways of the routing received by means of a command traceroute, but also and some analytical sizes, as that distribution of number of routers and numbers of independent systems on a way between two measuring points as it is shown in a Fig. 5.

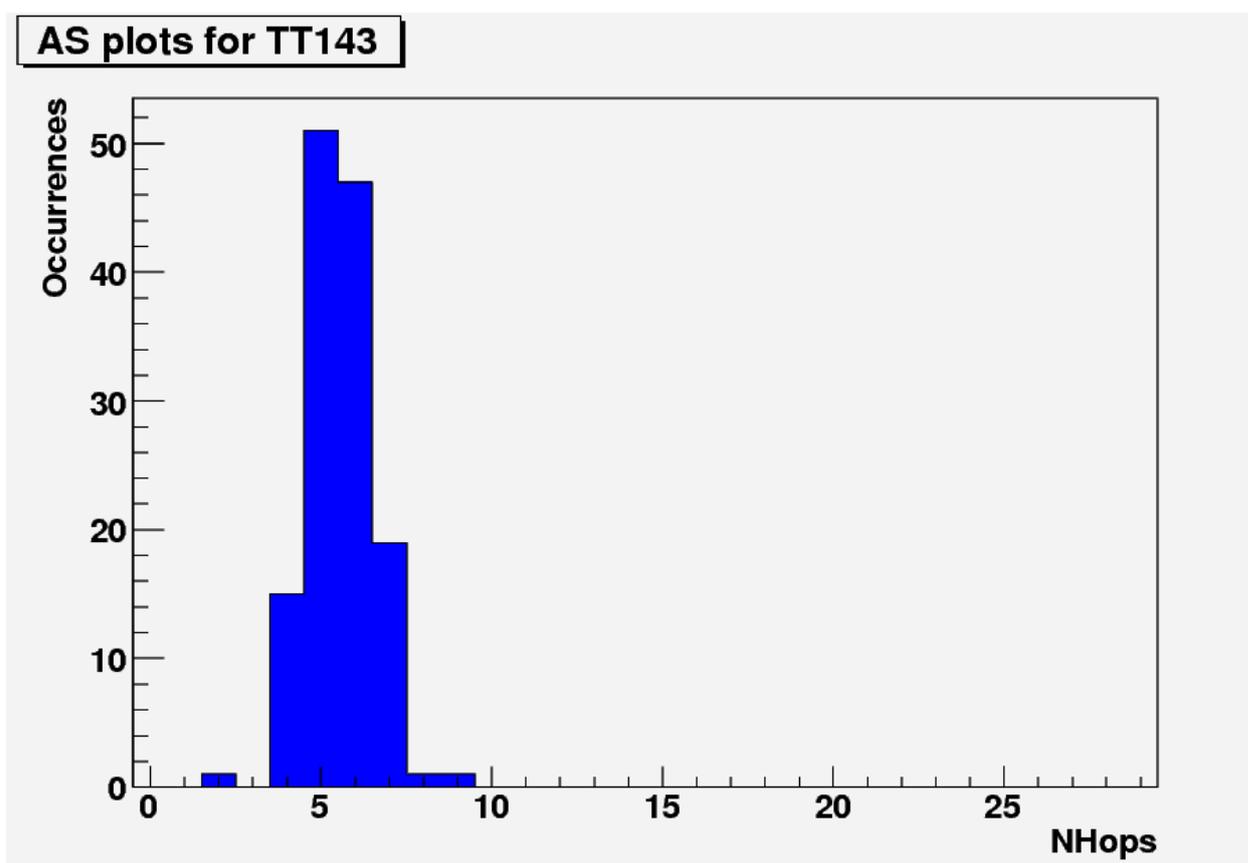

Fig. 5. AS plots for tt143

During operation of measuring system we had been eliminated some network incidents, most significant of them concerns to detection of inaccessibility of a root server of a domain name of the first level. System engineers of German national domain DE (a.nic.de), analyzed availability of the server by means of measuring system RIPE and had found out, that from a point in Samara and, accordingly, from all independent systems of network RBNet there is no opportunity to serve inquiries to the above-stated server.

**Summary and feature work**

In this work we present the simple method for estimation of available bandwidth. The detailed experiment data [2] allows us to modernize the expression for end-to-end available bandwidth and to derive the new formula where the well measured values are included.

We hope that our results will be incorporated in measurement mechanisms of RIPE test boxes [5] as well as in other measurement packets like [7] H.323 Beacon. Any VVoIP applications demand the knowledge of this value for tuning of transmitting process.

The additional experiments should be organized where the detailed comparison the results received with our method and with *iperf* utility [8]. It is interesting to investigate how the TCP throughput and available bandwidth depends from packet loss value.